\documentclass[conference]{IEEEtran}
\usepackage{blindtext, graphicx}
\usepackage{caption}
\usepackage{hyperref}

\usepackage{graphicx}
\usepackage{fixltx2e}
\usepackage{balance}
\usepackage{epstopdf}
\DeclareGraphicsExtensions{.eps}
\usepackage{amsmath}
\usepackage{float}
\usepackage{url}
\usepackage{balance}
\usepackage{fancyhdr}
\usepackage{cite}

\begin{document}

\markboth{Buch et al.}{Robust Real-time Impulsive RFI Mitigation for Radio Telescopes}
\pagestyle{fancy}
\fancyhf{}
\rhead{Submitted to the IETE Technical Review}
\title{Implementing and Characterizing Real-time Broadband RFI Excision for the GMRT Wideband Backend}
\author{\IEEEauthorblockN{Kaushal D. Buch, Kishor Naik, Swapnil Nalawade, Shruti Bhatporia,Yashwant Gupta, and Ajithkumar B.}
\IEEEauthorblockA{Digital Backend Group, Giant Metrewave Radio Telescope, \\National Center for Radio Astrophysics-Tata Institute of Fundamental Research, Pune, INDIA\\
Email: kdbuch@gmrt.ncra.tifr.res.in}
}
\maketitle
\begin{abstract}
The Giant Metrewave Radio Telescope (GMRT) is being upgraded to increase the receiver sensitivity. This makes the receiver more susceptible to man-made Radio Frequency Interference (RFI). To improve the receiver performance in presence of RFI, real-time RFI excision (filtering) is incorporated in the GMRT wideband backend (GWB). The RFI filtering system is implemented on FPGA and CPU-GPU platforms to detect and remove broadband and narrowband RFI. The RFI is detected using a threshold-based technique where the threshold is computed using Median Absolute Deviation (MAD) estimator. The filtering is carried out by replacing the RFI samples by either noise samples or constant value or threshold. This paper describes the status of the real-time broadband RFI excision system in the wideband receiver chain of the upgraded GMRT (uGMRT). The test methodology for carrying out various tests to demonstrate the performance of broadband RFI excision at the system level and on radio astronomical imaging experiments are also described.
\end{abstract}
\IEEEpeerreviewmaketitle

\section{Introduction}
GMRT is a array consisting of thirty, 45m diameter parabolic reflector antennas\cite{Swarup1991}. It is a passive radiometer for observing astrophysical phenomena occurring in the universe at radio frequencies from 150 MHz to 1450 MHz. As a part of the upgraded GMRT project, the bandwidth of the GMRT receiver is increased from 32 MHz to 400 MHz to improve the sensitivity \cite{gupta2017upgraded}. As a result, the more sensitive uGMRT receiver is now susceptible to manmade RFI. In order to achieve desired sensitivity in the presence of RFI, an RFI excision system is incorporated in the uGMRT receiver chain. This FPGA-based RFI excision system operates in real-time for filtering broadband RFI on time-series of individual antennas. The system is released as part of the GWB and is currently undergoing system-level testing for understanding the effects of broadband RFI filtering. \\

RFI excision techniques use a robust statistical estimator to estimate the dispersion of the signal and the filtering thresholds \cite{Fridman2008},\cite{Buch2016}. The signal values outside the threshold are blanked (replaced by zeros) or held at the threshold. Another approach uses non-normality detection to find and remove blocks of data that deviate from the normal distribution due to RFI \cite{series2013techniques}. The RFI excision technique used for uGMRT determines the filtering threshold through robust statistical estimate of dispersion using Median Absolute Deviation (MAD). The samples outside this threshold are replaced by a constant value or threshold or digital noise \cite{BuchR2016}. Early work in RFI mitigation at the Westerbork Synthesis Radio Telescope describes FPGA implementation of the algorithm and its effects on the cross-correlation power spectrum as described in \cite{baan2004radio}. The description in \cite{baan2004radio}, \cite{Fridman2008} provides the details of different estimators, their implementation, and tests. To the best of authors' knowledge, there has been no detailed study and a generic method for testing of the effects of real-time RFI excision on different measurement parameters of a radio telescope.  \\

This paper describes the broadband RFI excision technique and its implementation and testing on GWB. The contributions of this paper are: 

\begin{itemize}

\item{A comparative analysis of the signal with and without real-time broadband RFI filtering for the two different signal processing modes (correlator and beamformer) of the GWB.}

\item{Analysis and results from controlled tests were carried out using a programmable analog instrument developed at the GMRT to emulate desired type of RFI signal of a required strength.}

\item{The system performance in the presence of RFI by measuring improvement in the cross-correlation between antenna pairs and accessing the quality of astronomical imaging.}

\end{itemize}
The paper is organized as follows - Section II provides details of the technique of RFI excision, properties of RFI and the implementation of the RFI filtering system on GWB. Section III describes the test setup, system configuration and results from the various system tests and imaging experiments carried out for the broadband RFI excision. Section IV provides discussion and future work. \\

\section{Real-Time Broadband RFI excision in the GWB}
This section provides an overview of the broadband RFI, its sources around the GMRT and its effects on the astronomical signal. This is followed by a brief introduction to the GWB and the technique used for broadband RFI mitigation in the uGMRT.
\subsection{Broadband RFI at the GMRT}
Broadband RFI is a result of impulsive events in time-domain which results in an increase in the power across the spectrum. At GMRT, the sources of broadband RFI are sparking on high power lines or transformers, corona discharges and vehicular sparking in the vicinity of the telescope. RFI caused due to sparking on AC transmission lines is periodic over multiples of line frequency (50 Hz in India) - 10ms for a single-phase line and 3.3 ms for a 3-phase line. This type of RFI is observed as a group of strong impulses, 10 to 20 dB higher than the system noise and a duration of few tens of microseconds. Individual impulses are generally have a duration of tens of nanoseconds. The severity and strength of broadband RFI observed at the GMRT is stronger at lower radio frequencies and worsens during pre-monsoon and monsoon season. Strong broadband RFI superimposes on the Gaussian distributed astronomical signal and makes the distribution of the resulting signal otherwise wide-sense stationary, heavy-tailed\cite{Fridman2008}.
\subsection{Effects of broadband RFI}
Broadband RFI introduces a non-random component to the astronomical signal. As a result, the fluctuations in the measurement do not reduce upon temporal integration as per the radiometer equation, i.e.$\sqrt{BT}$, where B is the receiver bandwidth in Hz and T is the integration time in seconds.  The degradation in the signal-to-noise ratio (SNR) due to broadband RFI limits the detection of weak radio astronomical sources. Real-time RFI excision reduces the effects of broadband RFI significantly by removing it at an early stage (pre-detection or pre-correlation) in the signal processing chain.
\subsection{GWB}
GWB is a real-time signal processing system for the uGMRT receiver\cite{Reddy2017}. It is implemented using a combination of FPGAs and GPUs. This system digitizes the 60 baseband signal inputs(30 antennas dual polarization) each having a bandwidth of 400MHz. Real-time broadband RFI filtering is carried out for every input at the Nyquist rate (800 MHz) on FPGAs. The digitized signal is then fed to a \textit{compute} cluster which consists of high-end GPUs. GWB performs two main operations-correlation and beamforming. The correlator and beamformer outputs are temporally integrated to improve the SNR. The correlator provides auto and cross correlation spectrum between every pair of antennas every 671ms(minimum) whereas the beamformer provides four either incoherent or phased-array beam spectrum every 80$\mu s$(minimum). In the current GWB configuration, each FPGA-based hardware board can process up to a maximum of four inputs.
\subsection{RFI Excision technique}\label{rfiexctec}
Real-time broadband RFI excision process in the GWB is split in three parts-robust estimation, threshold detection, and filtering. Median Absolute Deviation ($MAD$) is used for robust estimation of dispersion of the data in presence of RFI (outliers). MAD ($D$) is then scaled to equivalent robust standard deviation (1.4826 times $D$ in case of normal distribution) for determining the detection thresholds 

\begin{align}
	{\tau_U} = {M(x)} + n \times (1.4826 \times  D) \nonumber \\
	{\tau_L} = {M(x)}  - n \times (1.4826 \times  D),
	\end{align}

around the median $M(x)$ of the data window.
The value $n$ decides the multiple of standard deviation that the threshold would be from the median of the data. Each input sample ($x_i$) in the data window is then compared with the detection thresholds ($\tau_u  and  \tau_l$). The output sample $Z_i$ is unchanged if the input sample is within the thresholds

\begin{align}
z_i  &= K  ; ~ for ~x_i \geq ~{\tau_u} ~or ~x_i \leq ~{\tau_l} \nonumber \\
&= x_i; ~for ~{\tau_l} < ~x_i < ~{\tau_u}
\end{align}

 whereas the samples outside the thresholds are replaced with constant value, digital noise \cite{buch2014variable} or threshold ($K$).
MAD estimator is robust up to 50\% RFI in a given data window. Thus, the typical duration of RFI and rate at which the signal is sampled determine the size of the data window for MAD estimation. Since real-time MAD computation over large window sizes is impractical to develop due to hardware resource requirements, an alternate technique called Median of MAD (MoM) \cite{BuchR2016} is introduced in the GWB. Instead of depending on one window for estimating the MAD value, MoM($M_m$) uses median($M$) of the $k$ MAD values
\begin{equation}
\label{4}
M_m = M(D_1,D_2,D_3,\dots,D_k)
\end{equation}
where ${D_1,D_2,...,D_k}$ are the MAD values for 'k' successive data windows. MoM based threshold estimation is useful in mitigating longer bursts of strong RFI. 

Currently, the real-time broadband RFI excision system in the GWB has two variants, MAD and MoM. MAD estimation is carried out on a window of 16384 samples and MoM is a median of 4096 MAD values where each MAD value is estimated over 4096 data samples. For a 1.25 ns sampling period in the GWB, the maximum duration of RFI that the RFI excision system can filter is 40$\mu s$  for MAD and 40$ms$ for MoM.\\
\section{Testing Broadband RFI Excision on GWB: Schemes And Results}
\begin{figure*}[!th]
\centering
\includegraphics[width=\textwidth]{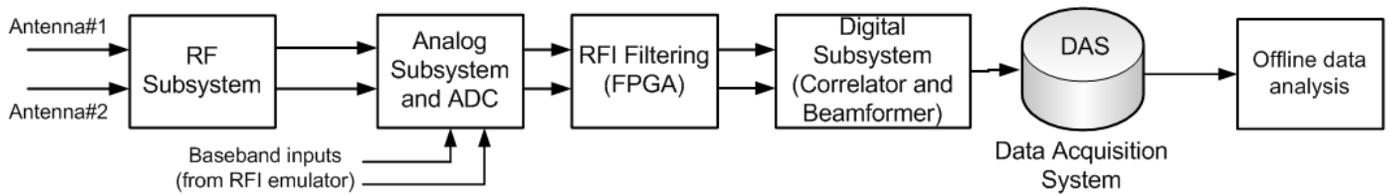}
\caption{Block diagram of the test setup for real-time RFI excision for the uGMRT}
\label{gwb_sys}
\end{figure*}

The radio frequency signal is received, processed and conditioned before digitization as shown in Figure \ref{gwb_sys}. The digitized signal is sent to the FPGA boards where the broadband RFI filtering is implemented in time-domain. The output of the FPGA boards is fed to a CPU-GPU cluster for correlation and beamforming operations. The output of this cluster is sent to a acquisition computer for recording the data. Analysis is carried out on recorded data.\\

To observe and analyze the RFI instances before and after RFI filtering, the beamformer output is preferred because of its relatively short integration time. The beamformer mode eases locating the occurrences of broadband RFI in a long stretch of data. Since this technique is implemented for the correlator mode of the GWB, the effect of filtering is checked on the correlator output as well. During some comparison tests, both, correlator and beamformer outputs are simultaneously recorded. The results described in this paper are for 2048 spectral channels, 1.3 ms (beamformer mode) and 671 ms (correlator mode) settings of the GWB. The temporal behavior of a single spectral channel is shown in all the results described in this paper. The time axis in all the plots represents the actual Indian Standard Time (IST) of the observations.\\

\subsection{Test schemes}
Broadband RFI filtering is implemented on FPGA board with four digital inputs. The two basic ways in which the tests were carried out used the digital copy of the signals inside the FPGA. 
\begin{itemize}
\item {The first option uses the unfiltered and filtered copies of two antennas to be sent to the digital signal processing subsystem for further processing. With this arrangement, the beamformer and correlator measurements can be recorded simultaneously.}
\item{ The second option uses a single antenna input copied to all the four inputs out of which three were filtered. This arrangement facilitates simultaneous analysis of two variable parameters of this filter-threshold value ($n$) and replacement option ($K$) for the same antenna input signal.}\\
\end{itemize}
Quantitative improvement in the signal-to-noise ratio (SNR) was carried out using the metric shown in\cite{Buch2016}. The improvement (dB) is given as $10\log_{10}(S_U/S_F)$  where $S_U$ and $S_F$ are the average mean-to-RMS ratios of the unfiltered and filtered signals respectively.
\subsection{Test results}
This section describes the results from the simultaneous comparison tests between - correlator and beamformer outputs, different replacement options and MAD and MoM. This is followed by the test results from the system-level off-source tests and imaging tests.

\subsubsection{Simultaneous comparison of correlator and beamformer outputs}
\begin{figure*}[!th]
\centering
\includegraphics[width=\textwidth]{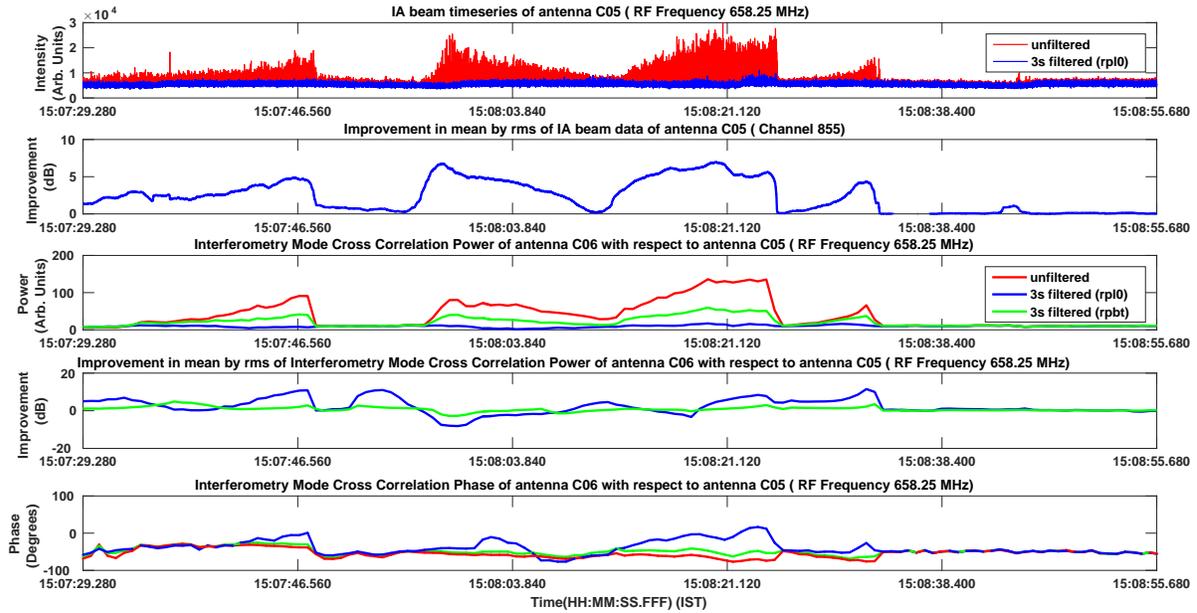}
\caption{Beamformer and correlator outputs for the comparison of different replacement options}
\label{punezoom}
\end{figure*}

The effect of RFI filtering using MAD at $3\sigma$ threshold and replacement by zero value for a single antenna can be seen in the first subplot of Figure \ref{punezoom}. The filtered (blue) and unfiltered (red) outputs in subplot 1 and the corresponding improvement metric in dB is shown in subplot 2. Simultaneous recording in the correlator mode is also carried out for collocated antennas in order to observe correlated RFI. It can be seen that the cross-correlation between the unfiltered signals shows an increase in the power level during instances of RFI (subplot 3). This increase in the power level is filtered as can be seen in the correlation between the filtered signals. A comparison shows that the filtering  is more for replacement by zero than by a non-zero threshold as seen in subplot 4. Subplot 5 shows the behaviour of the phase of the cross-correlation for unfiltered and filtered options. In subplot 5, the replacement by threshold maintains the phase of cross-correlation constant and has hence is better than replacement by zeros where the phase changes.\\
\subsubsection{Comparison of replacement options using beamformer output}
\begin{figure*}[!th]
\centering
\includegraphics[width=\textwidth]{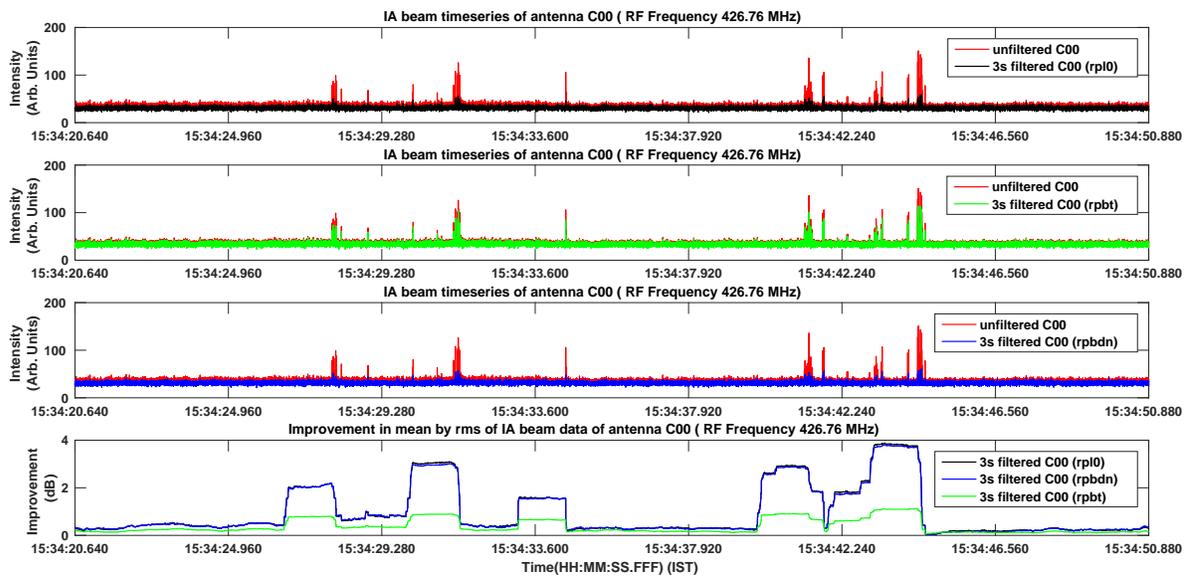}
\caption{Beamformer outputs for comparison of different replacement options using the 1:4 digital copy mode}
\label{replcompmad}
\end{figure*}

A simultaneous comparison of three replacement options (zero, digital noise, and threshold) is carried out on four simultaneous beamformer outputs (as described in Section III A). The comparison is shown in Figure \ref{replcompmad}. The first subplot shows comparison between unfiltered data (red) and replacement by zero (black). The second and third subplot show a comparison between unfiltered data (red) and replacement by threshold (green) and digital noise (blue) respectively. The fourth subplot shows Improvement metric for the three replacement schemes. Replacement by zero value provides the best improvement as the RFI instances are replaced by the lowest possible value. Replacement by digital noise provides improvement similar to that of replacement by zeros. Replacement by threshold does not provide good improvement as the RFI values are clipped at the thresholds which is a much higher value as compared to the other two filtering options. \\
\subsubsection{Comparison between MAD and MoM}
\begin{figure*}[!th]
\centering
\includegraphics[width=\textwidth]{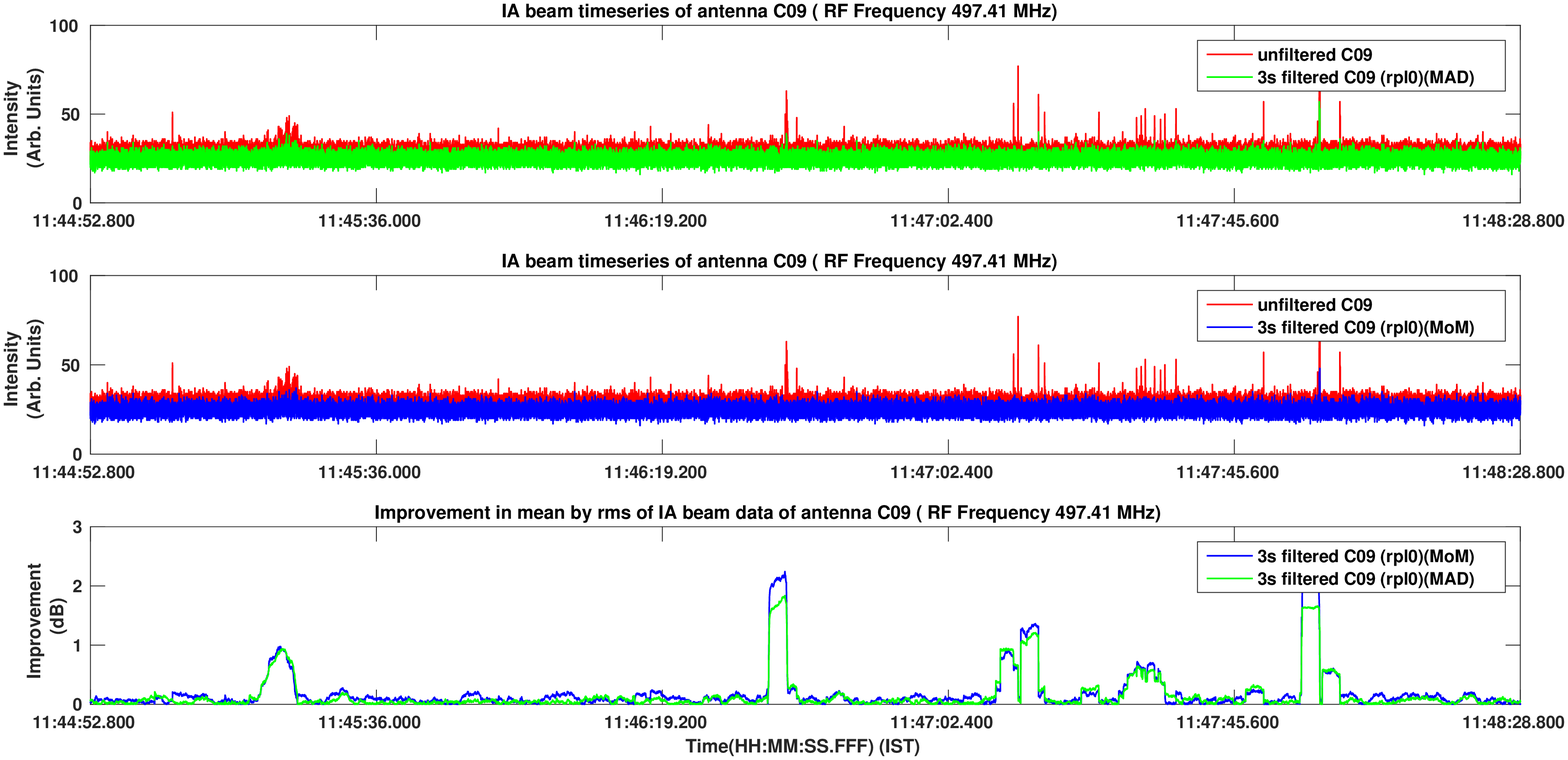}
\caption{Beamformer outputs from simultaneous comparison of MAD and MoM techniques}
\label{madmomant}
\end{figure*}

A comparison between the MAD and MoM techniques for real-time excision is required to demonstrate their performance for different RFI conditions. The test setup is similar to that described in Section III A. In Figure \ref{madmomant}, the first subplot shows an overlay of unfiltered (red) and filtered (MAD) (green) option. The second subplot shows that MoM (green) performs better over the MAD option. The third subplot is the comparison in the Improvement metric. It can be seen that in this case both techniques perform well except for one instance where the RFI is excised more strongly using the MoM technique. \\

\begin{figure*}[!th]
\centering
\includegraphics[width=\textwidth]{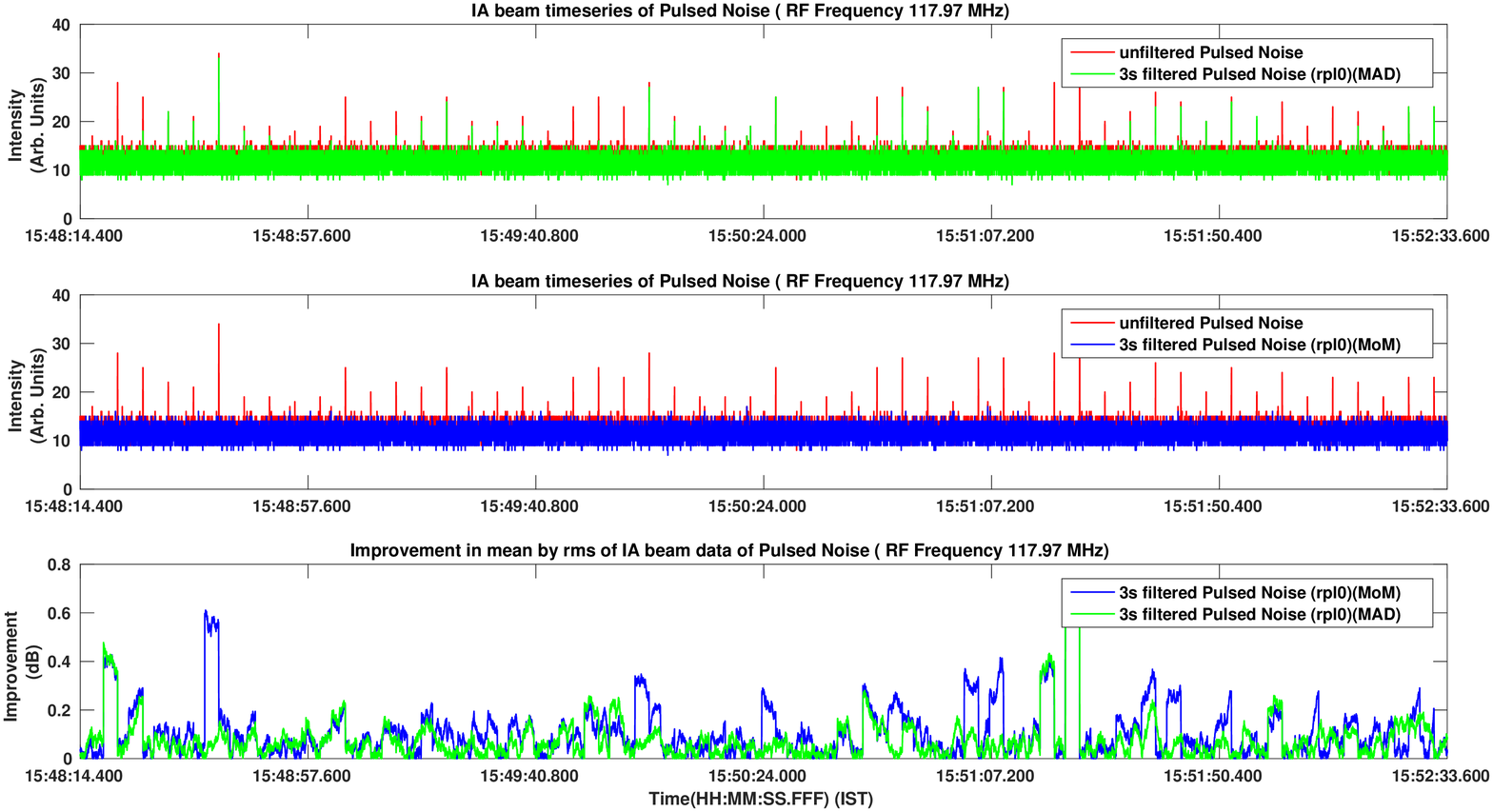}
\caption{Beamformer outputs from simultaneous comparison of MAD and MoM techniques for emulated RFI signal }
\label{madmomcompem}
\end{figure*}

A better comparison can be seen in Figure \ref{madmomcompem} where the same test setup is provided with analog inputs at base-band frequency (Figure \ref{gwb_sys}) through RFI emulator. RFI emulator is an analog instrument developed at the GMRT for emulating the behavior of different types of RFI, particularly the power-line RFI with a specific duty cycle. Broadband RFI behavior is emulated with an on-period 64us and total period of 2s for the RFI emulator. It can be seen from Figure \ref{madmomcompem} that the MAD technique is unable to detected many RFI instances (subplot 1) particularly those which persist for duration greater than 10$\mu s$. On the other hand, the MoM technique detects all the RFI instances. The improvement metric also shows better performance of MoM over MAD (subplot 3).
\subsubsection{Off-source tests}
\begin{figure*}[!th]
\centering
\includegraphics[width=\textwidth]{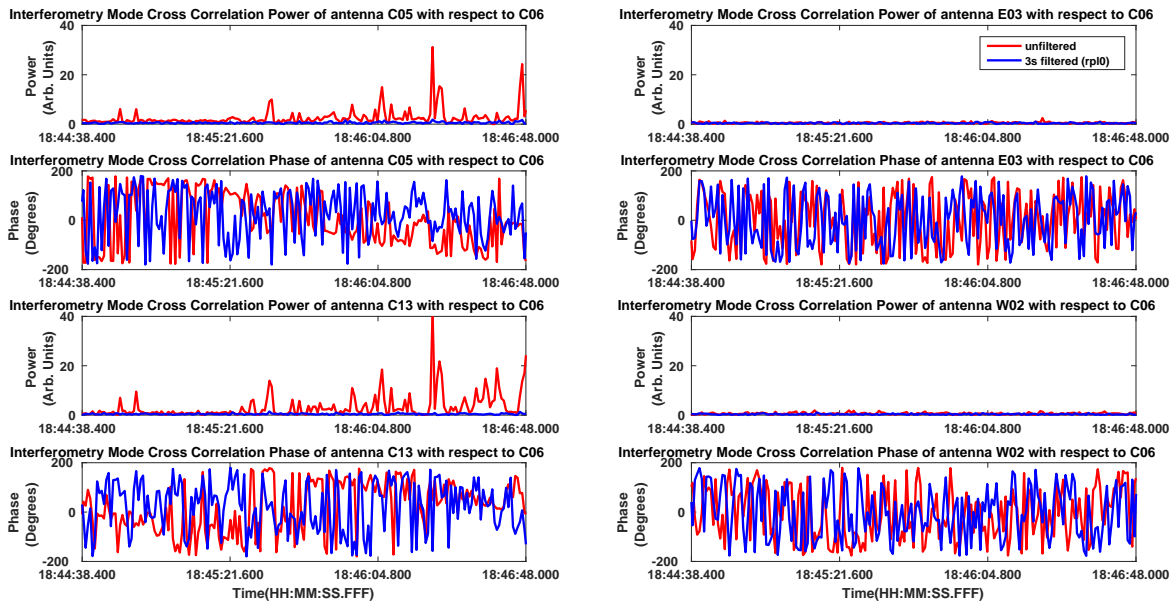}
\caption{Off-source tests showing the effect of real-time broadband RFI filtering}
\label{offsrc}
\end{figure*}

Since broadband RFI primarily occurs due to sparking on high-power transmission lines around the GMRT array, it is found to be correlated for the closely spaced antennas. As the spacing between the antennas increases, RFI becomes less correlated. This leads to a spurious increase in the correlation even when the antennas are not observing a radio source (i.e. off-source). In order to study the effect of real-time broadband RFI filtering, tests were carried out when the antennas were pointing 5 degrees off-source. Figure \ref{offsrc} shows single spectral channel magnitude and phase of the cross correlation for two short spacing antennas (left) and two long spacing antennas (right). Simultaneous observations were carried out to observe the effect with and without filtering in the 250-500 MHz band of uGMRT. For nearby antenna pairs, the effect of filtering on the magnitude of cross correlation(first and third subplots in the left panel) is significant at $2\sigma$ threshold. The post-filtering phase of the correlation(second and fourth subplots in the left panel) becomes random indicating the reduction in the correlated component in the signal. The antenna pairs separated by longer distances are not affected as seen in the subplots in the right panel.\\

\begin{figure*}
\centering
\includegraphics[width=\textwidth]{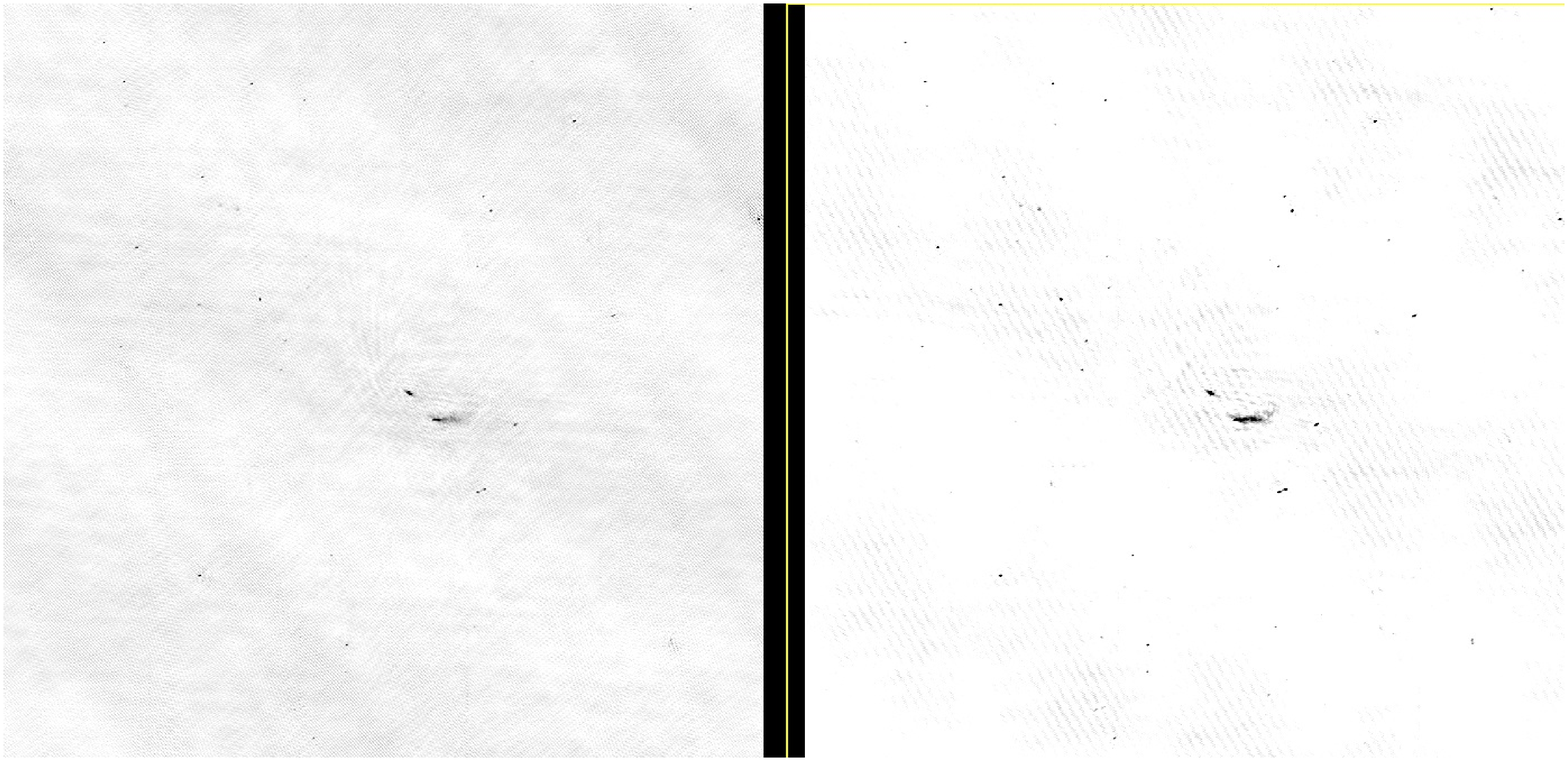}
\caption{Radio image from 16-antenna uGMRT observation in the 250-500 MHz band-one polarization with real-time broadband RFI filtering (right) and the other polarization without filtering(left) (Image Courtesy: Dharam Vir Lal)}
\label{rimage}
\end{figure*}

\subsubsection{uGMRT Test Observations}
A radio image generated from uGMRT observation is shown in Figure \ref{rimage}. The test setup had 16 antennas having 200MHz bandwidth in the 250-500 MHz RF band. One of the polarizations (left-circular) is processed through the RFI filter (left image) and the other polarization (right-circular) is processed without the filter (right image). The assumption is that broadband RFI is unpolarized. RFI filtering is carried out at $3\sigma$ threshold and the RFI is blanked (replaced by zeros). The image formed without the filter has artifacts and also has a  lesser contrast. This is expected as broadband RFI adds to the overall noise floor and hence results in reduced dynamic range of the radio image. It can be seen that the image obtained after RFI filtering is much cleaner and has improved contrast over the one without the filter. In this clean image, it is possible to distinguish a few background radio sources which were obscured due to the RFI in the image without RFI filtering. There is an overall improvement in the noise (RMS value) by a factor of two after RFI filtering in case of the image (Figure \ref{rimage}). \\

\section{Discussion and future work}
The paper described the implementation and testing of real-time broadband RFI excision technique for the uGMRT. Real-time implementation of different variants of broadband RFI filtering which are released for uGMRT system were described. The procedure and performance of various methods developed for testing the excision using antenna and emulator signals was described. The results show that the filter is able to significantly reduce the effects of RFI, leading to an improved receiver sensitivity. Real-time RFI excision enables better imaging of weak astronomical sources in presence of strong RFI.  Such tests are being repeated to observe the effects of filtering on astronomical imaging and to establish an optimal filtering threshold and replacement strategy.\\

Using the techniques for testing suggsted in this paper, a more detailed analysis of the effects of threshold and replacement options on the various observation parameters and astronomical imaging is being carried out. The techniques described here are generic and can be applied to any radio telescope or radiometer for testing and characterizing real-time RFI excision.

\section*{Acknowledgment}
The authors would like to thank the members of the Backend group and Operations group at the GMRT for their help and support. The authors would also like to thank Dr. Dharam Vir Lal and Sanjay Kudale of NCRA-TIFR for their help in planning the test observations, data analysis, and imaging.


\balance
\end{document}